\documentclass[a4paper]{spie}  

\usepackage[]{graphicx}
\usepackage{amssymb,amsmath,psfrag,url}

\newcommand{\Field}[1]{{\boldsymbol{#1}}}
\newcommand{\Kvec}[1]{{\vec{#1}}}

\title{EMF Simulations of Isolated and Periodic 3D Photomask Patterns}

\author{
Sven Burger\supit{\,ab}, 
Lin Zschiedrich\supit{\,ab},
Frank Schmidt\supit{\,ab}, 
Roderick K\"ohle\supit{\,c}, \\
Bernd K\"uchler\supit{\,d}, 
Christoph N\"olscher\supit{\,d}
\skiplinehalf
\supit{a}
Zuse Institute Berlin,
Takustra{\ss}e 7,
D\,--\,14\,195 Berlin,
Germany\\
DFG Forschungszentrum {\sc Matheon},
Stra{\ss}e des 17.\,Juni 136, 
D\,--\,10\,623 Berlin,
Germany
\smallskip\\
\supit{b}
JCMwave GmbH,
Haarer Stra{\ss}e 14a,
D\,--\,85\,640 Putzbrunn, 
Germany
\smallskip\\
\supit{c}
Qimonda AG,
Advanced Technology Software\\
Am Campeon 1-12, D\,--\,85\,579 M\"unchen, Germany
\smallskip\\
\supit{d}
Qimonda Dresden GmbH \& Co.OHG, QD P LM F\\
K\"onigsbr\"ucker Stra{\ss}e 180,
D\,--\,01\,099 Dresden,
Germany
}

\authorinfo{
Corresponding author: S. Burger\\
URL: http://www.zib.de/Numerik/NanoOptics/\\
Email: burger@zib.de
}

\begin{document} 
  \maketitle 
\noindent
Copyright 2007  Society of Photo-Optical Instrumentation Engineers.\\
This paper will be published in Proc.~SPIE Vol. {\bf 6730}
(2007),  
({\it Photomask Technology, R. J. Naber, H. Kawahira, Eds.})
and is made available 
as an electronic preprint with permission of SPIE. 
One print or electronic copy may be made for personal use only. 
Systematic or multiple reproduction, distribution to multiple 
locations via electronic or other means, duplication of any 
material in this paper for a fee or for commercial purposes, 
or modification of the content of the paper are prohibited.
\begin{abstract}

We present rigorous 3D EMF simulations of isolated 
features on photomasks using a newly developed finite-element 
method. 
We report on the current status of the finite-element 
solver JCMsuite, incorporating higher-order edge elements, 
adaptive refinement methods, and fast 
solution algorithms.
We demonstrate that rigorous and accurate results on light 
scattering off isolated features can be achived at relatively 
low computational cost, compared to the standard approach 
of simulations on large-pitch, periodic computational domains. 

\end{abstract}



\section{Introduction}

Rigorous electromagnetic field (EMF) simulations have become an important 
tool for mask design in low $k_1$ applications. 
As has been shown in previous works finite-element 
simulations are superior in terms of simulation 
accuracy, convergence rate and computation speed 
when compared to other presently used EMF simulation 
methods~\cite{Burger2005bacus,Burger2006c}.
We address 3D simulation tasks occuring in microlithography
by using the frequency-domain FEM solver {\it JCMsuite}. 
This solver has been successfully applied to a wide 
range of 3D electromagnetic field computations including
microlithography~\cite{Burger2005bacus,Burger2006c,Koehle2007a,Burger2007om},
left-handed metamaterials in the optical 
regime~\cite{Linden2006a}, and
photonic crystals~\cite{Burger2006b}.
The solver has also been used for pattern reconstruction in 
EUV scatterometry~\cite{Pomplun2006bacus,Tezuka2007a}, and it has been benchmarked 
to other methods in typical DUV lithography
 projects~\cite{Burger2005bacus,Burger2006c}.

In this paper we report on the current status of the finite-element 
solver {\it JCMsuite}, and we present simulation results of light transition through 
isolated and periodic features on photomasks.

\section{Background}

The finite-element package {\it JCMsuite} allows to simulate a variety 
of electromagnetic problems. It incorporates a scattering solver ({\it JCMharmony}), 
a propagating mode solver ({\it JCMmode}) and a resonance solver ({\it JCMresonance}).
The scattering, eigenmode and resonance problems can be formulated 
on 1D, 2D and 3D computational domains. 
Admissible geometries can consist of 
periodic or isolated patterns, or a mixture of both.
Further, solvers for 
problems posed on cylindrically symmetric geometries are implemented. 

\begin{figure}[htb]
\centering
\psfrag{-inf}{\sffamily $-\infty$}
\psfrag{+inf}{\sffamily $+\infty$}
\psfrag{CDx}{\sffamily $\mbox{CD}_{\mbox{x}}$}
\psfrag{px}{\sffamily $\mbox{p}_{\mbox{x}}$}
\includegraphics[height=0.3\textwidth]{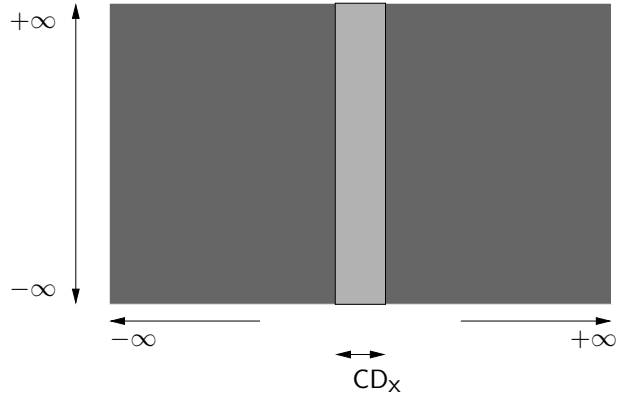}
\caption{
Schematics of a mask with an isolated line: cross-section in a $x$-$y$-plane. 
The line with a critical dimension of $CD_x$  is depicted
in light grey. 
The pattern extends towards infinity in positive and negative $x$- and $y$-directions.
}
\label{schema_geo_5}
\end{figure}

\begin{figure}[htb]
\centering
\psfrag{-inf}{\sffamily $-\infty$}
\psfrag{+inf}{\sffamily $+\infty$}
\psfrag{CDx}{\sffamily $\mbox{CD}_{\mbox{x}}$}
\psfrag{p_x}{\sffamily $\mbox{p}_{\mbox{x}}$}
\psfrag{w_x}{\sffamily $\mbox{w}_{\mbox{x}}$}
\psfrag{d2}{\sffamily $\mbox{d}_{\mbox{Cr,top}}$}
\psfrag{d1}{\sffamily $\mbox{d}_{\mbox{Cr,bottom}}$}
\psfrag{substrate}{\sffamily substrate}
\psfrag{superspace}{\sffamily superspace (air)}
\psfrag{computational_domain}{\sffamily computational domain}
\includegraphics[height=0.3\textwidth]{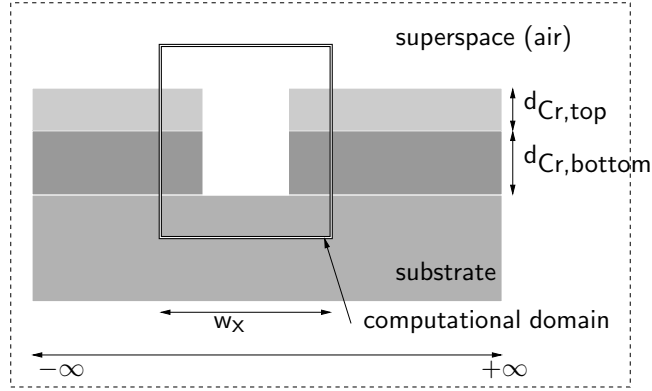}
\caption{
Schematics of an isolated line mask: cross-section in the $x$-$z$-plane. 
The material stack on a Quartz (SiO$_2$) substrate consists of 
two layers containing chromium. 
The computational domain is indicated. Please note that the computational domain is 
significantly smaller than the computational domain for a periodic line mask with  
high $p_x/CD_x$-ratio (see Fig.~\ref{schema_geo_7}).
}
\label{schema_geo_8}
\end{figure}

\begin{figure}[htb]
\centering
\psfrag{-inf}{\sffamily $-\infty$}
\psfrag{+inf}{\sffamily $+\infty$}
\psfrag{CDx}{\sffamily $\mbox{CD}_{\mbox{x}}$}
\psfrag{px}{\sffamily $\mbox{p}_{\mbox{x}}$}
\includegraphics[height=0.3\textwidth]{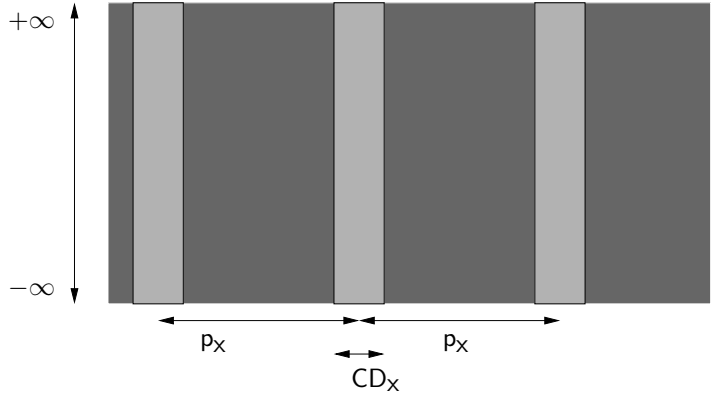}
\caption{
Schematics of a periodic line mask: cross-section in a $x$-$y$-plane. 
Lines with a critical dimension of $CD_x$  are depicted
in light grey. The pattern is periodic in $x$-direction with a pitch of $p_x$.
}
\label{schema_geo_4}
\end{figure}

\begin{figure}[htb]
\centering
\psfrag{-inf}{\sffamily $-\infty$}
\psfrag{+inf}{\sffamily $+\infty$}
\psfrag{CDx}{\sffamily $\mbox{CD}_{\mbox{x}}$}
\psfrag{p_x}{\sffamily $\mbox{p}_{\mbox{x}}$}
\psfrag{d2}{\sffamily $\mbox{d}_{\mbox{Cr,top}}$}
\psfrag{d1}{\sffamily $\mbox{d}_{\mbox{Cr,bottom}}$}
\psfrag{substrate}{\sffamily substrate}
\psfrag{superspace}{\sffamily superspace (air)}
\psfrag{computational_domain}{\sffamily computational domain}
\includegraphics[height=0.3\textwidth]{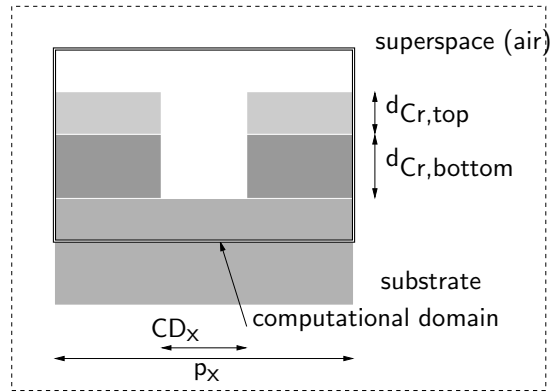}
\caption{
Schematics of a periodic line mask: cross-section in the $x$-$z$-plane. 
The material stack on a Quartz (SiO$_2$) substrate consists of 
two layers containing chromium. 
The computational domain is indicated. 
}
\label{schema_geo_7}
\end{figure}

\begin{figure}[htb]
\centering
\psfrag{Layout}{\sffamily Geometry layout}
\psfrag{Triangulation}{\sffamily Mesh parameters}
\psfrag{Materials}{\sffamily Material parameters}
\psfrag{Mesh}{\sffamily Mesh}
\psfrag{Sources}{\sffamily Sources parameters}
\psfrag{Project}{\sffamily Project parameters}
\psfrag{Results}{\sffamily Results}
\psfrag{Projection}{\sffamily Projection}
\psfrag{Interface}{\sffamily MATLAB Interface: Automatic input generation,
parameter scans, data analysis}
\includegraphics[width=0.85\textwidth]{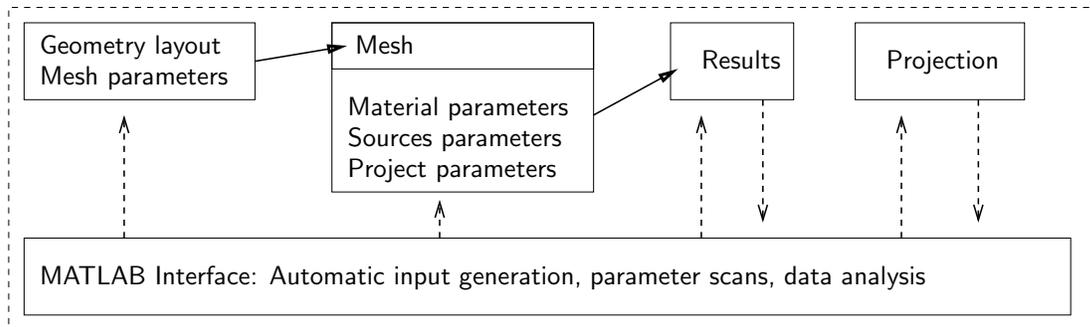}
\caption{
Schematics of the simulation flow of the FEM solver 
{\it JCMsuite}.
}
\label{schema_simulation_flow}
\end{figure}

\begin{table}[h]
\begin{center}
\begin{tabular}{|l|l|l|l|l|l|l|}
\hline
layout setup& isolated& \multicolumn{2}{c|}{periodic} \\ 
\hline
  & set 1&  set 2&  set 3\\ 
\hline 
\hline 
$\mbox{CD}_{\mbox{x}}$ [nm]  & \multicolumn{3}{l|}{260}\\
$\mbox{d}_{\mbox{Cr,top}}$ [nm] & \multicolumn{3}{l|} {18}\\
$\mbox{d}_{\mbox{Cr,bottom}}$ [nm] & \multicolumn{3}{l|}{55}\\
\hline 
$p_x$ & -- & $6\times \lambda_0$ : $4\times \lambda_0$ : $206\times \lambda_0$  & $1~\mu$m : 40\,nm : $20~\mu$m \\ 
$w_x$ & 400\,nm       & --   & -- \\ 
\hline 
\hline 
$\lambda_0$ & \multicolumn{3}{l|} {193.0\,nm , coherent illumination, $\sigma =0$ } \\ 
$\varepsilon_{r \mbox{air}}$ & \multicolumn{3}{l|} {1.0}\\
$\varepsilon_{r \mbox{Cr,top}}$ & \multicolumn{3}{l|} {$(1.965 + 1.201i)^2$}\\
$\varepsilon_{r \mbox{Cr,bottom}}$ & \multicolumn{3}{l|}{$(1.477 + 1.762i)^2$}\\
$\varepsilon_{r \mbox{SiO}_2}$ & \multicolumn{3}{l|} {$(1.564)^2$}\\
\hline 
\hline 
N.A. & \multicolumn{3}{l|} {1.0 } \\ 
Magnification & \multicolumn{3}{l|} {4 X } \\ 
Rel.~threshold & \multicolumn{3}{l|} {0.69 } \\ 
\hline 
\end{tabular} 
\caption{Parameter settings for the isolated (data set 1) and periodic (data sets 2 and 3) line mask 
simulations: Mask geometry, material parameters, illumination, imaging parameters. 
The width of the computational window in the isolated case is $w_x=400\,$nm (set 1), the pitch is varied in steps of 
$4\times \lambda_0$, resp.~in steps of 40\,nm in the periodic case (set 2, resp.~3).  
The relative permittivity, $\varepsilon_r$, is the square of the complex 
index of refraction, $\varepsilon_r=n^2=(n_r+i k)^2$.
}
\label{table_linemask}
\end{center}
\end{table}

In this paper we concentrate on 
light-scattering off 2D and 3D photomask patterns (lines and contact holes). 
The patterns of interest are isolated in  the $x-$ and $y-$directions and are 
enclosed by  homogeneous 
substrate and superstrate (typically air) which are infinite 
in the $-z$-, resp.~$+z$-direction. 
Cross sections through these patterns are schematically shown 
in Figures~\ref{schema_geo_5} and~\ref{schema_geo_8}. 
We compare the results obtained from calculations on isolated computational domains 
to results from calculation on periodic computational domains with a large pitch.
Cross sections through the periodic  patterns are schematically shown 
in Figures~\ref{schema_geo_4} and~\ref{schema_geo_7}. 

Light propagation in the investigated systems is governed by Maxwell's equations
where  vanishing densities of free charges and currents are assumed~\cite{Wong2005a}. 
The dielectric coefficient $\varepsilon(\vec{x})$ and the permeability 
$\mu(\vec{x})$ of the considered photomasks are complex and in the case of periodic 
mask layouts periodic, 
$\varepsilon \left(\vec{x}\right)  =  \varepsilon \left(\vec{x}+\vec{a} \right)$, 
$\mu \left(\vec{x} \right)  =  \mu \left(\vec{x}+\vec{a} \right)$.
Here $\vec{a}$ is any elementary vector of the periodic lattice.  
For given primitive lattice vectors 
$\vec{a}_{1}$ and $\vec{a}_{2}$ a periodic elementary cell 
$\Omega\subset\mathbb R^{3}$ can be defined as
$\Omega = \left\{\vec{x} \in \mathbb R^{2}\,|\,
x=\alpha_{1}\vec{a}_1+\alpha_{2}\vec{a}_2;
0\leq\alpha_{1},\alpha_{2}<1
\right\}
\times [z_{sub},z_{sup}]$.
In the isolated case the computational domain is choosen such that in the 
exterior domain only homogeneous or waveguide-like structures are present
(see Fig.~\ref{schema_geo_8}).

A time-harmonic ansatz with frequency $\omega$ and magnetic field 
$\Field{H}(\vec{x},t)=e^{-i\omega t}\Field{H}(\vec{x})$ leads to
the following equations for $\Field{H}(\vec{x})$:
\begin{itemize}
\item
The wave equation and the divergence condition for the magnetic field:
\begin{eqnarray}
\label{waveequationH}
\nabla\times\frac{1}{\varepsilon(\vec{x})}\,\nabla\times\Field{H}(\vec{x})
- \omega^2 \mu(\vec{x})\Field{H}(\vec{x}) &=& 0,
\qquad\vec{x}\in\Omega,\\
\label{divconditionH}
\nabla\cdot\mu(\vec{x})\Field{H}(\vec{x}) &=& 0,
\qquad\vec{x}\in\Omega .
\end{eqnarray}
\item
In the case of an isolated pattern:
Transparent boundary conditions at all boundaries of the computational domain, 
$\partial\Omega$,
where $\Field{H}^{in}$ is the incident magnetic field (plane waves 
in this case), and $\vec{n}$ is the normal vector on $\partial\Omega$:
\begin{equation}
\label{tbcH}
	\left(
        \frac{1}{\varepsilon(\vec{x})}\nabla \times (\Field{H} - 
        \Field{H}^{in})
	\right)
	\times \vec{n} = DtN(\Field{H} - 
        \Field{H}^{in}), \qquad \vec{x}\in \partial\Omega.
\end{equation}
The $DtN$ operator (Dirichlet-to-Neumann) is  realized with 
an adaptive PML method~\cite{Zschiedrich03,Zschiedrich2006a}. 
This is a generalized formulation of Sommerfeld's radiation condition.

\item 
In the case of a $x-y$-periodic pattern:
Transparent boundary conditions at the boundaries to the 
substrate (at $z_{sub}$) and superstrate (at $z_{sup}$), $\partial\Omega$,
according to Equation~\ref{tbcH},
and periodic boundary conditions for the transverse boundaries, $\partial\Omega$,
governed by Bloch's theorem~\cite{Sakoda2001a}:
\begin{equation}
\label{bloch}
\Field{H}(\vec{x}) = e^{i \Kvec{k}\cdot\vec{x}} \Field{u}(\vec{x}), \qquad
\Field{u}(\vec{x})=\Field{u}(\vec{x}+\vec{a}),
\end{equation}
where the Bloch wavevector $\Kvec{k}\in\mathbb{R}^3$ is defined by the
incoming plane wave $\Field{H}^{in}$.

\end{itemize}

Similar equations are found for the electric field 
$\Field{E}(\vec{x},t)=e^{-i\omega t}\Field{E}(\vec{x})$;
these are treated accordingly.
The finite-element method solves Eqs.~(\ref{waveequationH}) -- (\ref{bloch})
in their weak form, i.e., in an integral representation. 

The finite-element methods consists of the following steps:
\begin{itemize}
\item
The computational domain is discretized with simple geometrical patches,
{\it JCMsuite} uses linear (1D), triangular (2D) and tetrahedral or prismatoidal 
(3D) patches. 
The use of prismatoidal patches is well suited for layered geometries, as in 
photomask simulations. 
\item
The function spaces in the integral representation of Maxwell's equations 
are discretized using Nedelec's edge elements, 
which are vectorial functions of polynomial order 
defined on the simple geometrical patches~\cite{Monk2003a}. 
In the current implementation, {\it JCMsuite} uses polynomials of 
first ($1^{st}$) to ninth ($9^{th}$) order. 
In a nutshell, FEM can be explained as expanding the field 
corresponding to the exact solution of Equation~(\ref{waveequationH}) in the 
basis given by these elements.
\item
This expansion leads to  a large sparse matrix equation (algebraic problem).
To solve the algebraic problem on a standard workstation 
linear algebra decomposition techniques (LU-factorization, e.g.,
package PARDISO~\cite{PARDISO}, which was used in the simulations of 
Chapters~\ref{2dchapter},~\ref{3dchapter})
are used. 
In cases with either large computational domains or high accuracy 
demands, also domain decomposition methods~\cite{Zschiedrich2005b} 
are used and allow to handle problems with very large numbers of unknowns.
\end{itemize}

For details on the weak formulation, 
the choice of Bloch-periodic functional spaces,
the FEM discretization, and the implementation of the adaptive PML
method in {\it JCMsuite}
we refer to previous works~\cite{Zschiedrich03,Zschiedrich2006a}.
In future implementations performance will further be increased 
by using curvilinear elements, general domain-decomposition techniques  and 
$hp$-adaptive methods. 

\begin{figure}[h]
\centering
\psfrag{(a)}{\sffamily (a)}
\psfrag{(b)}{\sffamily (b)}
\psfrag{(c)}{\sffamily (c)}
(a)
\includegraphics[width=0.45\textwidth]{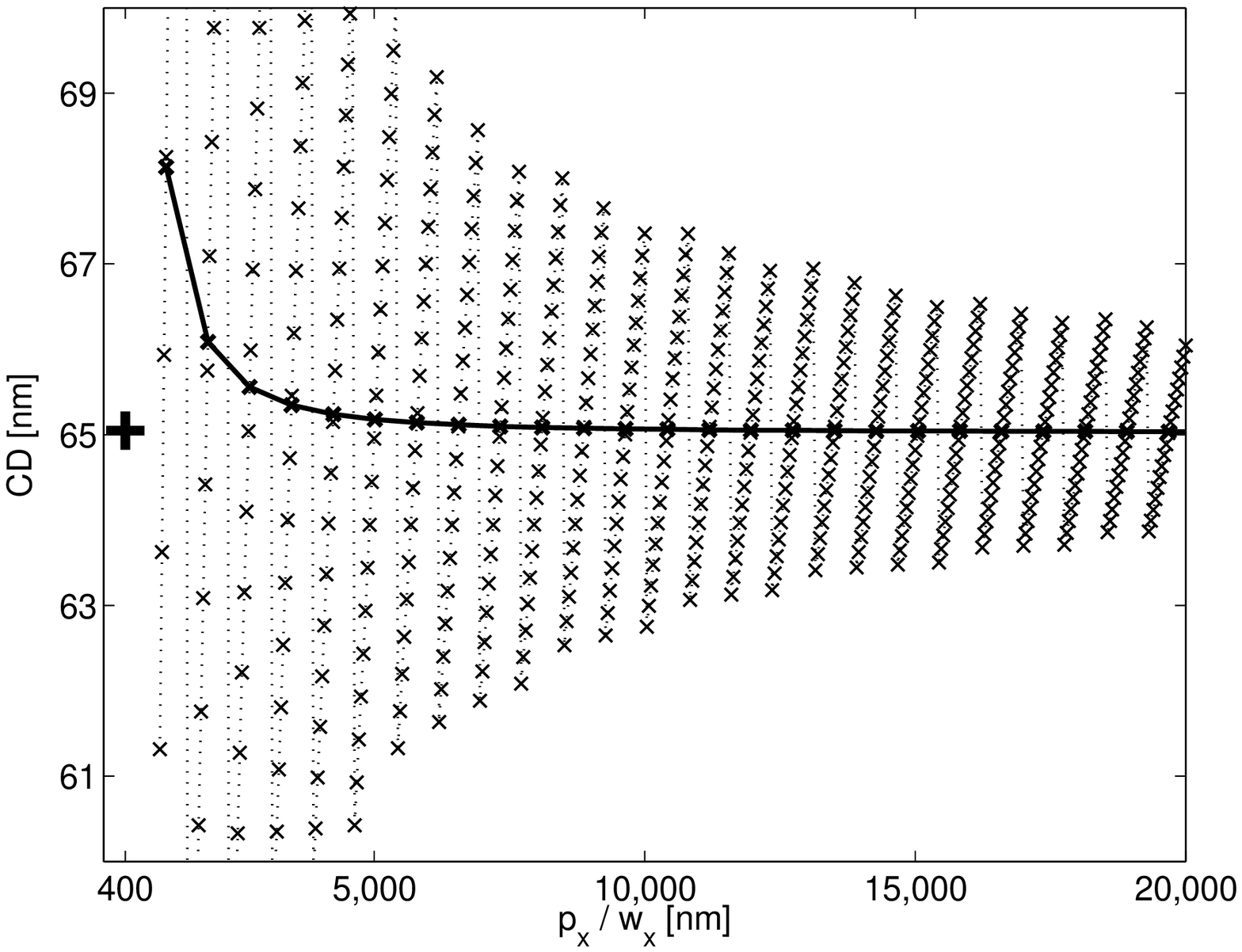}
(b)
\includegraphics[width=0.45\textwidth]{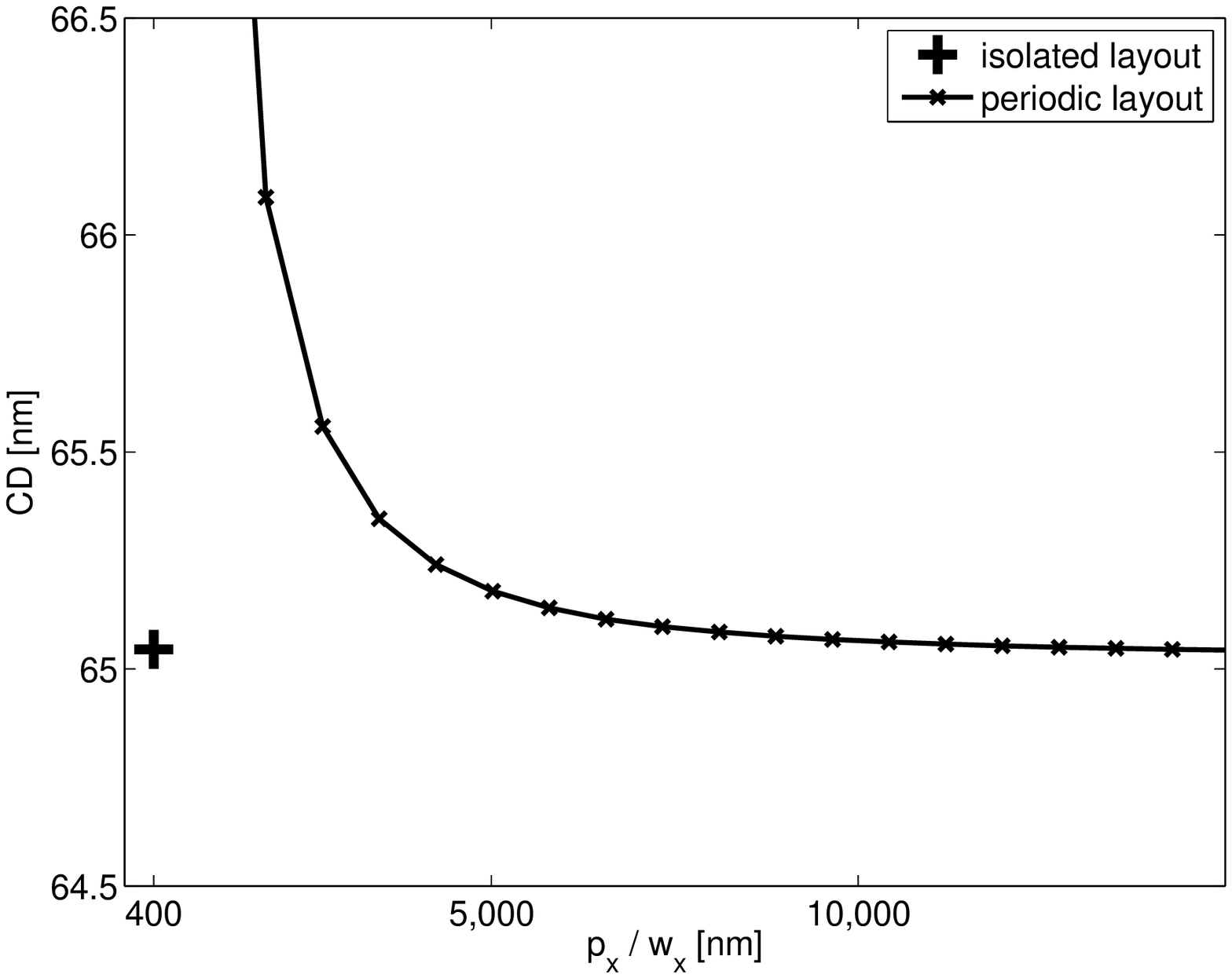}
\caption{
(a) Dependence of aerial image CD on computational domain width $w_x$, resp.~pitch $p_x$ 
for isolated and periodic layouts according to Table~\ref{table_linemask}. The single 
data point ({\bf$+$}) corresponds to the isolated layout and solver definitions (data set 1), 
the data points on the dotted/solid line correspond to the periodic layout (data set 2/3).
The strong variation of CD with pitch for the points on the dotted line is not observed when 
the CD is probed at specific pitch values only (solid line).  
(b) Detailed view of the same data as in (a), (data sets 1 and 3).}
\label{pitch_convergence}
\end{figure}

\section{Rigorous simulations and 
large-pitch periodic-domain simulations of an isolated line}
\label{2dchapter}

We apply the light-scattering solver of the programme package
{\it JCMsuite} in order to simulate the lithographic projection of an isolated line. 

The setup is as schematically shown in Figures~\ref{schema_geo_5} and~\ref{schema_geo_8}, with 
geometrical, material, illumination and imaging parameters as in Table~\ref{table_linemask} (data
set~1). 
We compare the obtained aerial image to results obtained with a setup of periodic lines of otherwise the same parameters, 
with a large ratio of pitch to linewidth (see 
Figures~\ref{schema_geo_4},~\ref{schema_geo_7} and Table~\ref{table_linemask}, data sets 2, 3).

Figure~\ref{schema_simulation_flow} shows the simulation flow of the 
finite-element software:
\begin{itemize}
\item
The geometry of the computational domain is described in a polygonal 
format  corresponding to Fig.~\ref{schema_geo_8} with  material
attributions and statements about the computational domain boundaries 
(periodic and transparent boundaries in this case).
The translational vectors of the periodic pattern ($\vec{a}_1, \vec{a}_2$)
are identified automatically from the layout, optimized settings 
of the perfectly matched layers (PML) are found automatically with an 
adaptive method~\cite{} (adaptive PML, aPML).
Further, parameters specifying a maximum patch size of the 
finite-element mesh and further meshing properties can be set.
From these geometry parameters and mesh parameters the  mesh is 
generated automatically. 
\item
Material parameters (complex permittivity and permeability tensors) 
can be specified as piecewise 
constant functions and/or (using dynamically loaded libraries)  
as functions of arbitrary spatial dependence. 
In the presented example, the piecewise constant, isotropic settings given 
in Table~\ref{table_linemask} are used. The relative permeability is $\mu_r=1$
for all used materials.
\item
Light sources can be defined as predefined functions (plane waves, Gaussian 
beams, point sources) or as arbitrary functions using dynamically loaded libraries. 
{\it JCMsuite} allows to generate solutions to several independent source terms 
in parallel, efficiently re-using the inverted system matrix. 
For the simulations in this chapter, without loss of generality, a coherent, polarized point source 
is used (wavevector $\vec{k}=(0,0,k_z)$). 
For the simulations in Chapter~\ref{3dchapter}, an unpolarized, incoherent source is used.
\item
The main project definitions in this case are the accuracy settings (mesh 
refinement, PML refinement, finite-element degree) and the definitions of 
postprocesses.
Upon execution the {\it JCMsolve} computes the full field distribution over the entire 
computational domain. 
Through internal or external post-processes, the quantities of interest can be derived from the 
field. E.g., the complex far field coefficients can be attained 
by Fourier integration / transformation and by the 
evaluation of the Rayleigh-Sommerfeld diffraction formula~\cite{Zschiedrich2007waves}, 
an aerial image is calculated, 
or the field distribution is exported to graphics format for 
visualisation and analysis.
\item
Interfaces to scripting languages like Python or MATLAB can be used for 
performing automatic parameter scans and data analysis.
In the presented example, the interface for the generation of 
the input files and for the execution of a loop over the 
geometrical parameters given in Tables~\ref{table_linemask},~\ref{table_3dmask} 
has been realized in 
MATLAB.
\end{itemize}

\begin{figure}[htb]
\centering
\psfrag{CDx}{\sffamily $\mbox{CD}_{\mbox{x}}$}
\psfrag{CDy}{\sffamily $\mbox{CD}_{\mbox{y}}$}
\psfrag{wx}{\sffamily $\mbox{w}_{\mbox{x}}$}
\psfrag{wy}{\sffamily $\mbox{w}_{\mbox{y}}$}
\psfrag{computational_domain}{\sffamily computational domain}
\includegraphics[height=0.3\textwidth]{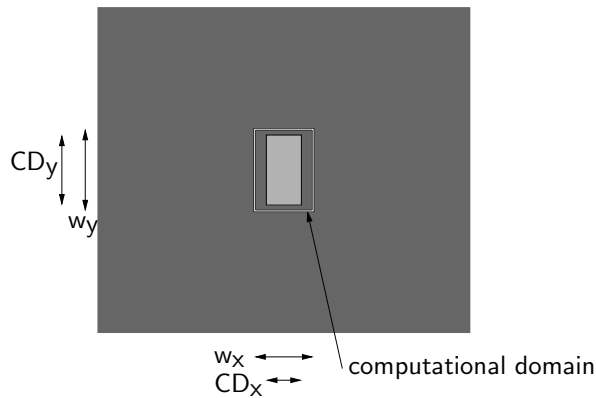}
\caption{
Schematics of a mask with an isolated contact hole: cross-section in an $x$-$y$-plane. 
The contact hole with a critical dimensions in $x$-, resp.~$y$-direction of $CD_x$ / $CD_y$    is depicted
in light grey. 
The pattern extends towards infinity in positive and negative $x$- and $y$-directions.
A cross-section through the computational domain is indicated.  
}
\label{schema_geo_10}
\end{figure}

\begin{figure}[htb]
\centering
\psfrag{CDx}{\sffamily $\mbox{CD}_{\mbox{x}}$}
\psfrag{CDy}{\sffamily $\mbox{CD}_{\mbox{y}}$}
\psfrag{px}{\sffamily $\mbox{p}_{\mbox{x}}$}
\psfrag{py}{\sffamily $\mbox{p}_{\mbox{y}}$}
\psfrag{computational_domain}{\sffamily computational domain}
\includegraphics[height=0.3\textwidth]{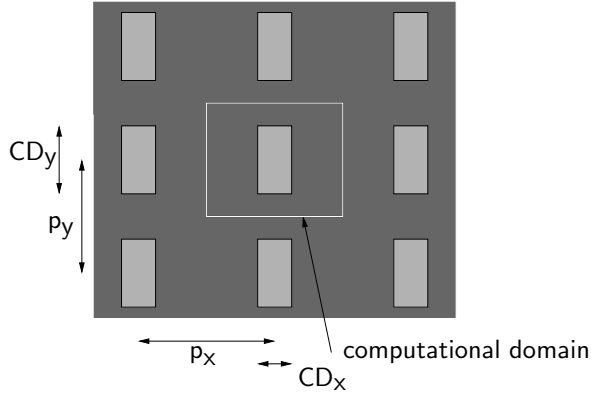}
\caption{
Schematics of a 2D-periodic array of contact holes: cross-section in a $x$-$y$-plane. 
Holes with critical dimensions of $CD_x$, $CD_y$  are depicted
in light grey. The pattern is periodic in $x/y$-direction with a pitch of $p_x/p_y$.
}
\label{schema_geo_9}
\end{figure}

\begin{table}[h]
\begin{center}
\begin{tabular}{|l|l|l|}
\hline
layout setup& isolated&periodic \\ 
\hline
  & set 1&  set 2\\ 
\hline 
\hline 
$\mbox{CD}_{\mbox{x}}$ [nm]  & \multicolumn{2}{l|}{240}\\
$\mbox{CD}_{\mbox{y}}$ [nm]  & \multicolumn{2}{l|}{300}\\
$\mbox{d}_{\mbox{Cr,top}}$ [nm] & \multicolumn{2}{l|} {18}\\
$\mbox{d}_{\mbox{Cr,bottom}}$ [nm] & \multicolumn{2}{l|}{55}\\
\hline 
$w_x$ & 400\,nm       & --    \\ 
$w_y$ & 400\,nm       & --    \\ 
$p_x$ & -- & $600\,$nm : 40\,nm : $4~\mu$m \\ 
$p_y$ & -- & $p_y = p_x$ \\ 
\hline 
\hline 
$\lambda_0$ & \multicolumn{2}{l|} {193.0\,nm , incoherent illumination, $\sigma =0$} \\ 
$\varepsilon_{r \mbox{air}}$ & \multicolumn{2}{l|} {1.0}\\
$\varepsilon_{r \mbox{Cr,top}}$ & \multicolumn{2}{l|} {$(1.965 + 1.201i)^2$}\\
$\varepsilon_{r \mbox{Cr,bottom}}$ & \multicolumn{2}{l|}{$(1.477 + 1.762i)^2$}\\
$\varepsilon_{r \mbox{SiO}_2}$ & \multicolumn{2}{l|} {$(1.564)^2$}\\
\hline 
\end{tabular} 
\caption{Parameter settings for the isolated (data set 1) and periodic (data set 2) contact hole 
mask simulations: Mask geometry, material parameters, illumination, imaging parameters.
The width of the computational window in the isolated case is $w_x=400\,$nm $\times$ 
$w_y=400\,$nm (set 1).  
The pitch is varied in steps of 40\,nm in the periodic case (set 2).  
The source is an unpolarized, single-point source. 
}
\label{table_3dmask}
\end{center}
\end{table}

From the obtained aerial image intensity distribution we decude 
a mask CD of the printed feature. 
For the specific parameter setting as given in Table~\ref{table_linemask} and from 
a simulation on the isolated computational domain we 
obtain a well converged value of the CD of approximately 65\,nm. 
When we instead simulate the light distribution using {\it periodic}  computational domains 
with large pitches we obtain some dependence of the CD on the pitch of the chosen 
computational domain. 
As expected, with increasing ratio of pitch to linewidth, 
the CD resulting from the aerial image 
at focal position with a
fixed threshold (0.69 of the maximum intensity in this case) 
is converging to a fixed value for 
pitches well above 5 microns. 
This behavior is depicted in Figure~\ref{pitch_convergence}. 
Please note that due to across-pitch imaging effects the resulting 
CD variation is rather large 
for a quasi continuously varied pitch (see, e.g.~\cite{Smith2003a}), 
see the dotted line in 
Figure~\ref{pitch_convergence}\,(a).
A smoother dependence is observed when the pitch is probed at multiples of wavelength 
times imaging magnification, see solid line in Figure~\ref{pitch_convergence}\,(a).
However, also with this method, the waver scale CD error at a pitch-to-linewidth ratio of 10:1
is still of the order of 1\,nm.
Therefore, for accurate results on isolated features using a periodic model with large 
pitch-to-linewidth ratio, very large computational domains have to be chosen.

This demonstrates the advantages using the non-periodic model 
according to Fig.~\ref{schema_geo_8} and
 Equation~\ref{tbcH}.
Using this model, the computational domain can be chosen just slightly larger than the 
feature of interest. 
The simulated aerial image still coincides with the aerial image from the periodic model with 
largest investigated pitch. 
This makes it possible to accurately compute images of isolated features 
with low computational effort (see 
singular data point for the isolated case in Figure~\ref{pitch_convergence} a,b).

\section{3D isolated contact hole - near field convergence}
\label{3dchapter}

\begin{figure}[htb]
\centering
\includegraphics[height=0.5\textwidth]{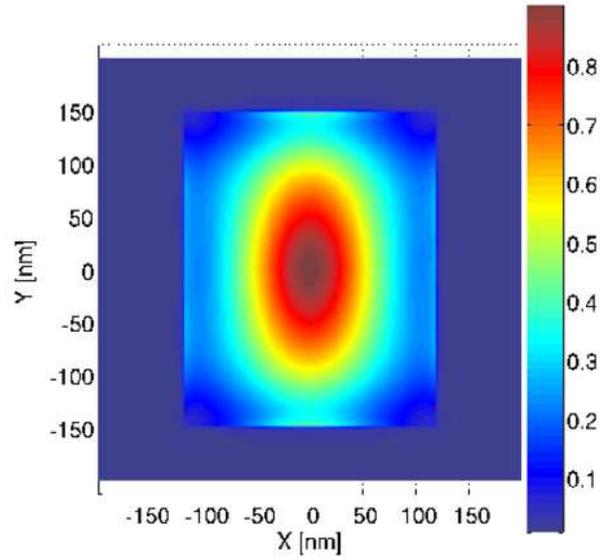}
\caption{
2D cross section through the 3D near field intensity distribution at the 
upper chromium layer. 
The colorbar on the right indicates the values for the field intensity  $I=I(x,y,z_0)$
(arbitrary units).
}
\label{cross_section_2d}
\end{figure}

\begin{figure}[h!]
\centering
\includegraphics[height=0.5\textwidth]{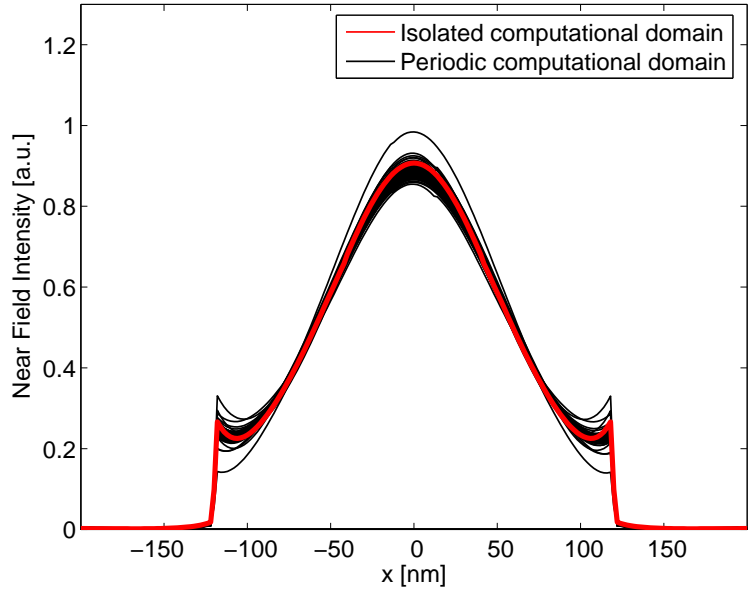}
\caption{
1D cross section through the 3D near field intensity distribution. 
The red solid line corresponds to the isolated case 
(Table~\ref{table_3dmask}, data set 1).
Black lines correspond to the periodic comparison case 
with different pitches as given in Table~\ref{table_3dmask}, data set 2.
}
\label{cross_section_1d}
\end{figure}

\begin{figure}[htb]
\centering
\includegraphics[height=0.5\textwidth]{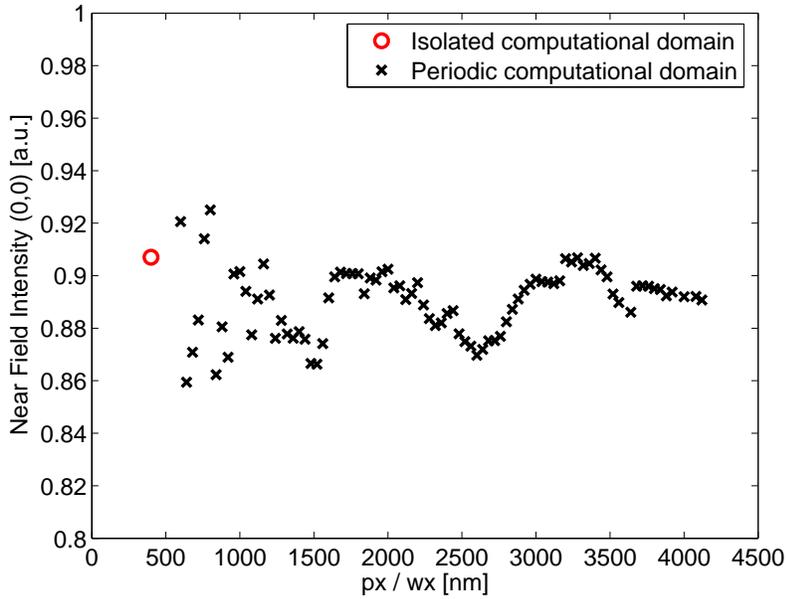}
\caption{
Field intensity of the data from Figure~\ref{cross_section_1d} at 
$x=0$. The data point for the simulation on an isolated domain is 
indicated by a red circle, the data points for simulations on periodic domains with 
varied pitches $p_x$ and $p_y=p_x$ is indicated by crosses. 
}
\label{cross_section_0d}
\end{figure}

\begin{figure}[htb]
\centering
\includegraphics[height=0.5\textwidth]{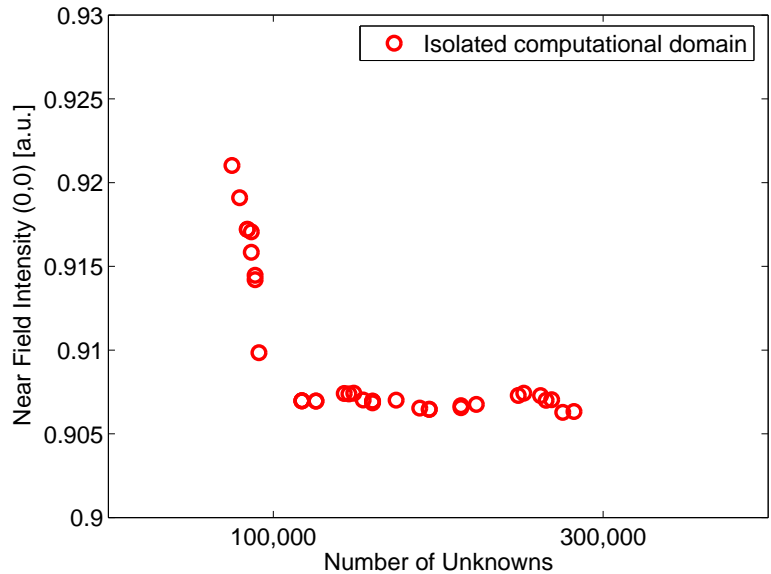}
\caption{
Convergence of the near field intensity ($I(\vec{r}_0)$ as 
function of the number of unknowns in the simulation).
}
\label{conv_iso}
\end{figure}

\begin{figure}[htb]
\centering
\includegraphics[height=0.5\textwidth]{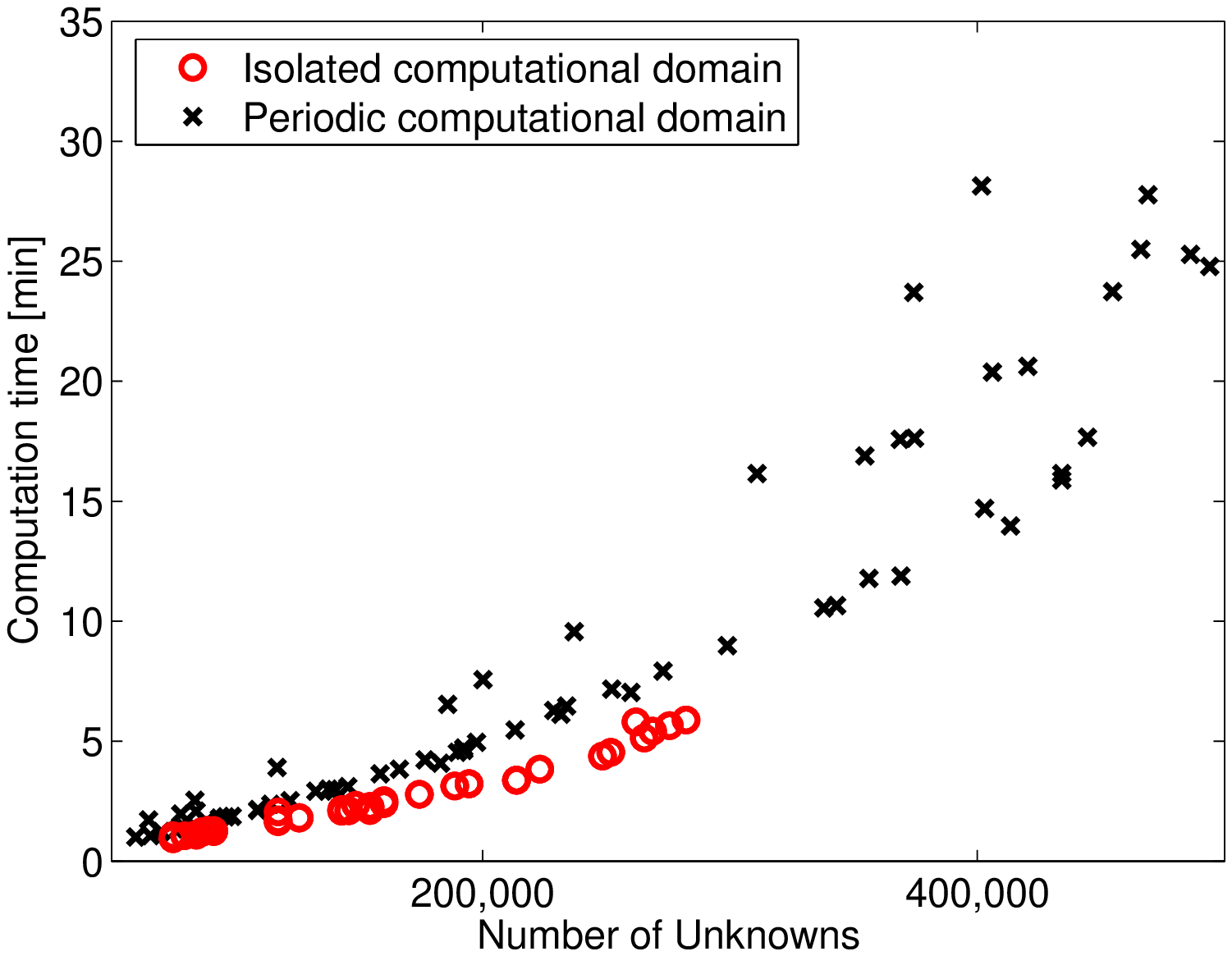}
\caption{
Computational cost for simulations on isolated domains (red circles)
and on periodic domains (black crosses). 
CPU time in minutes as function of the number of unknowns in the 
simulation. 
Please note that a simulation with $N$ unknowns on an isolated computational domain 
has in general a significantly better accuracy than a simulation with a similar number 
of unknowns on a (large-pitch) periodic computational domain (compare also 
Figures~\ref{cross_section_0d} and~\ref{conv_iso}). I.e., with a computation time of 
2\,min on an isolated domain a more accurate result is achieved than with a computation time 
of 20\,min on  a large-pitch periodic computational domain. 
}
\label{tcpu_n}
\end{figure}

In this chapter we investigate the application of our method to the simulation 
of light scattering off a 3D isolated contact hole. We concentrate on the convergence 
of the near field and -- as in Chapter~\ref{2dchapter} -- 
on the comparison to the large-pitch periodic case. 
Figure~\ref{schema_geo_10} shows a schematics of the geometrical layout:
A contact hole with square shape of width $CD_x$ and length $CD_y$ (in mask scale 
dimensions) spans in the $x$-$y$-plane. A cross-section in the $x$-$z$-plane
corresponds the linemask case, as depicted in Figure~\ref{schema_geo_8}. 

Figure~\ref{cross_section_2d} shows a $x-y$-cross section through the obtained 
near field intensity distribution. The cross section is taken at constant $z$ position 
of $z_0=-d_{Cr,top}/2$, where $z=0$ corresponds to the upper edge of the upper chromium layer. 
The illumination in this case is an unpolarized, incoherent point-source at 
perpendicular incidence from the substrate.
For the finite-element expansion and for the realisation of 
the PML boundary condition, fourth-order finite-elements have been used. 
The calculation has been performed on a standard 64-bit-CPU workstation with extended 
memory and several dual core processors (AMD Opteron).
The computation time for this problem with about 100,000 unknowns 
is of the order of one minute. 
The convergence of the intensity $I(0,0,z_0)$ is plotted in Figure~\ref{conv_iso}. 
Here, fourth order finite elements are used, and 
simulations with different numbers of unknowns in the FEM expansion are reached 
by different spatial discretizations of the computational domain. 
Figure~\ref{tcpu_n} shows the computational effort in minutes of total computation time 
in dependence of numbers of unknowns in the FEM expansion of the near field solution, $N$. 
The values give only an estimate of the computational costs as upon the computations, 
the workstation has been used 
for other tasks in parallel. 
Typical memory (RAM) usage for the FEM computation of the solutions corresponding to the 
data points of this Figure is between 1 and 15~GigaByte. 

Figure~\ref{cross_section_1d} shows a 1D cross section through the intensity $I(x,y,z)$ of the
 near field solution  
at constant $z_0=-d_{Cr,top}/2$ and $y=0$. 
Additional to the result obtained on the isolated periodic domain, results obtained on 
periodic domains are plotted in this figure. 
Due to typical computational resources limitations of full 3D computations the pitch 
in this case can not be choosen very large. Therefore the variation of the obtained field 
distributions in the investigated parameter regime is still of significant magnitude. 

Figure~\ref{cross_section_0d} shows the intensity values corresponding to the data 
from Figure~\ref{cross_section_1d} at $x=0$. 
As can be seen from the figure and as in the line mask case, a rather large computational 
domain has to be choosen when one tries to compute an accurate near field solution of an isolated 
contact hole on a periodic computational domain. 
The computational costs of the isolated and periodic computations as depicted in 
Figures~\ref{cross_section_0d} and~\ref{conv_iso} are shown in Figure~\ref{tcpu_n}.
As can be seen from the Figures, an accurate solution where the field intensity at a specified 
position is converged with an error of lower than 0.5\%
can be reached with about 100,000 unknowns on the isolated computational domain. 
In contrast, with a large-pitch periodic model, the pitches in $x$- and $y$-direction have 
to be chosen very large to ensure that the model-induced error is neglectible. 
This leads to a very high number of unknowns in the FEM problem and consequently to 
a highly increased computational effort.
Therefore, and especially for more complex isolated problems under investigation, a large-pitch periodic 
model is not a feasible option. 
This also holds for other discretization methods than FEM using periodic models 
for isolated problems. 
For the demonstrated 3D test problem the advantage of using isolated computational 
domains with transparent boundary conditions on all boundaries is more than one order of 
magnitude in computation time. 
With the implementation of this method rigorous and accurate investigation of scattering from isolated features 
has become straight-forward. 

\section{Conclusions}
\label{conclusions}
We have performed rigorous 3D FEM simulations of light transition through 
isolated features on linemasks and contact hole photomasks using the finite-element 
programme package {\it JCMsuite}. 
We have investigated the convergence behavior of the solutions and we have 
shown that we achieve results at high numerical accuracy. 
We have compared the results to calculations on periodic, large-pitch 
domains and found a performance advantage (cpu time) of at least one order of magnitude for 
calculations on isolated computational domains. 
Our results show that rigorous 3D simulations of isolated features on masks or wafers or in other 
settings can well be handled at high accuracy and relatively low computational cost. 


\bibliography{/home/numerik/bzfburge/texte/biblios/phcbibli,/home/numerik/bzfburge/texte/biblios/group,/home/numerik/bzfburge/texte/biblios/lithography}   
\bibliographystyle{spiebib}   

\end{document}